\documentclass[9pt,twocolumn,twoside]{opticaArxiv}

\usepackage{siunitx}
\usepackage[utf8]{inputenc}

\title{High-harmonic generation in periodically poled waveguides}

\author[1,*]  {Daniel~D.~Hickstein}
\author[1]    {David~R.~Carlson}
\author[1]    {Abijith~Kowligy}
\author[2]    {Matt~Kirchner}
\author[2]    {Scott~R.~Domingue}
\author[1]    {Nima~Nader}
\author[1]    {Henry~Timmers}
\author[1,3]  {Alex~Lind}
\author[4]    {Gabriel~G.~Ycas}
\author[2,3,5]{Margaret~M.~Murnane}
\author[2,3,5]{Henry~C.~Kapteyn}
\author[1]    {Scott~B.~Papp}
\author[1,3]  {Scott~A.~Diddams}

\affil[1]{Time and Frequency Division, National Institute of Standards and Technology, Boulder, Colorado 80305, U.S.A.}
\affil[2]{Kapteyn--Murnane Laboratories Inc., Boulder, CO 80301, U.S.A.}
\affil[3]{Department of Physics, University of Colorado, Boulder, Colorado, 80309, U.S.A.}
\affil[4]{Applied Physics Division, National Institute of Standards and Technology, Boulder, Colorado 80305, U.S.A.}
\affil[5]{JILA - University of Colorado Department of Physics and NIST, Boulder, CO 80305, U.S.A}
\affil[*]{Corresponding author: danhickstein@gmail.com}

\dates{\today}
\ociscodes{(190.4160) Multiharmonic generation; (140.3490) Lasers; (190.7110) Ultrafast nonlinear optics}
\doi{}

\begin{abstract}
Optical waveguides made from periodically poled materials provide high confinement of light and enable the generation of new wavelengths via quasi-phase-matching, making them a key platform for nonlinear optics and photonics. However, such devices are not typically employed for high-harmonic generation. Here, using 200-fs, 10-nJ-level pulses of 4100~nm light at 1~MHz, we generate high harmonics up to the 13$\mathrm{^{th}}$ harmonic (315~nm) in a chirped, periodically poled lithium niobate (PPLN) waveguide. Total conversion efficiencies into the visible--ultraviolet region are as high as 10~percent. We find that the output spectrum depends on the waveguide poling period, indicating that quasi-phase-matching plays a significant role. In the future, such periodically poled waveguides may enable compact sources of ultrashort pulses at high repetition rates and provide new methods of probing the electronic structure of solid-state materials.
\end{abstract}

\begin{document}

\maketitle

%%% SECTION: Introduction
\section{Introduction}
The process of high harmonic generation (HHG) allows intense, long-wavelength lasers to generate bursts of high energy photons, sometimes with pulse durations below 100~attoseconds~\cite{zhao_tailoring_2012}. Such ultrashort pulses have found numerous applications, including imaging nanometer-scale structures \cite{sandberg_lensless_2007, gardner_subwavelength_2017, zurch_real-time_2014}, observing the motions of molecules on the sub-femtosecond timescale \cite{leone_attosecond_2016, zhou_probing_2012}, and probing ultrafast charge- and spin-dynamics in materials \cite{kfir_nanoscale_2017, tao_direct_2016, chen_distinguishing_2017}. HHG is typically accomplished in atomic gases, but the recent observation of HHG in solids \cite{ghimire_observation_2011, hohenleutner_real-time_2015, gholam-mirzaei_high_2017, han_high-harmonic_2016, kim_generation_2017, luu_extreme_2015, ndabashimiye_solid-state_2016, schubert_sub-cycle_2014, yoshikawa_high-harmonic_2017} has sparked interest that solid-state materials could serve as a compact source of ultrashort pulses, and that the harmonic generation process itself may provide a versatile method for mapping the electronic structure of materials \cite{vampa_all-optical_2015, lanin_mapping_2017, you_anisotropic_2017, tancogne-dejean_impact_2017}. Although initial experiments utilized mid-infrared lasers to generate harmonics below the bandgap of the material,  more recent experiments have generated harmonics far above the bandgap and into the EUV region, with photon energies higher than 30~eV \cite{kim_generation_2017, luu_extreme_2015, ndabashimiye_solid-state_2016}.

% Figure: overview
\begin{figure}
	\includegraphics[width=\linewidth]{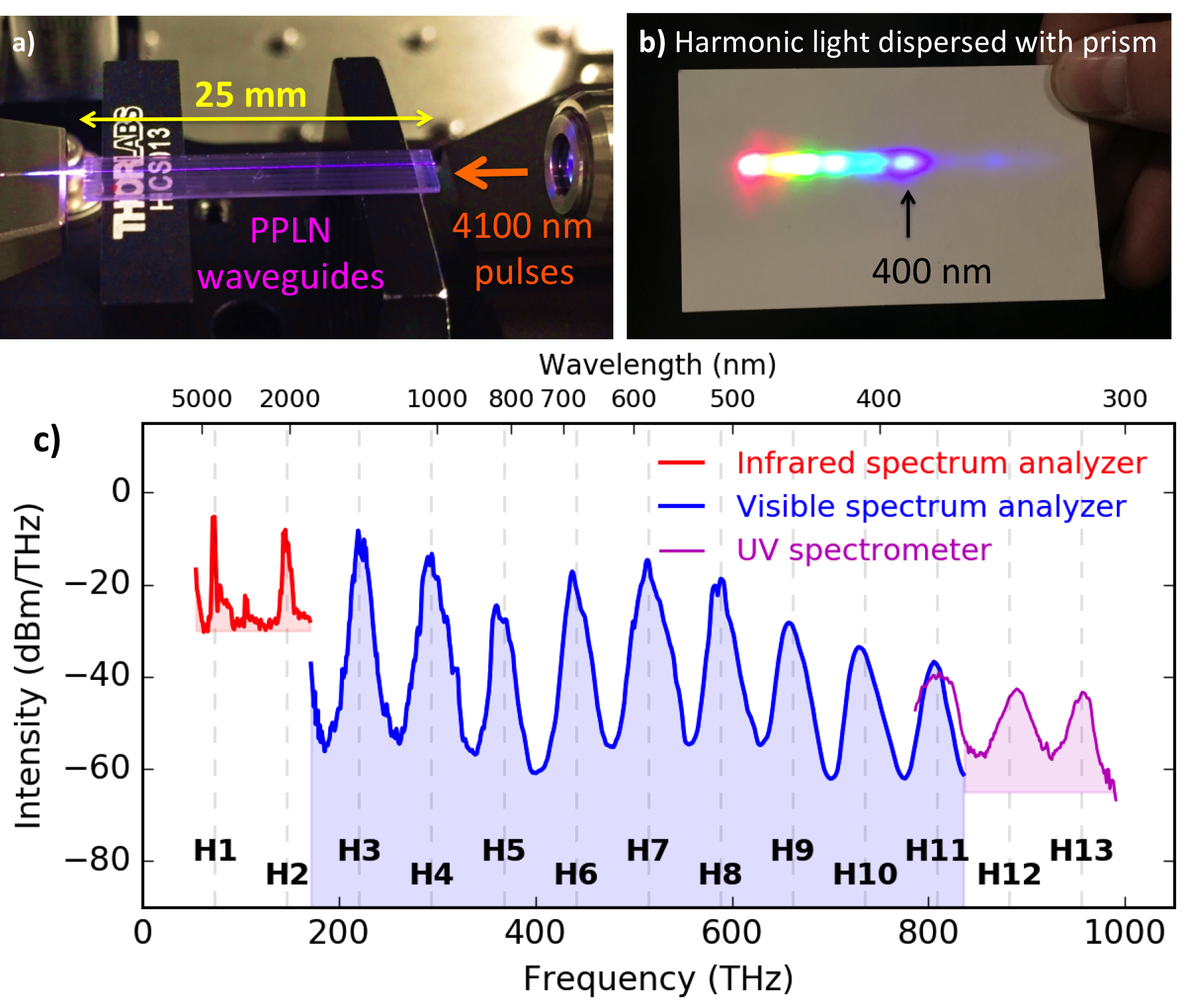}
	\caption{\label{fig:overview} a) When pumped with mid-infrared light, the periodically poled lithium niobate (PPLN) waveguide glows with visible light. b) The generated light, as dispersed with a prism and observed on white paper, reveals harmonic peaks across the visible and ultraviolet regions. c) When pumped with 20~nJ, $\sim$200~fs pulses of 4100~nm light, the output spectrum consists of harmonics up to the 13$^\mathrm{th}$ order (H13 at 315~nm).}
\end{figure}

While phase-matching has been extensively studied for HHG in gases \cite{rundquist_phase-matched_1998, tamaki_highly_1999, durfee_phase_1999, zhang_quasi-phase-matching_2007, cohen_grating-assisted_2007}, phase-matching of HHG in solids is relatively unexplored. In general, HHG in solids is not phase-matched, and this phase-mismatch limits the conversion efficiency of the process~\cite{ghimire_generation_2012}. Phase matching of HHG is notoriously difficult, since most materials exhibit a large difference in the refractive index between the long-wavelength fundamental light and the short-wavelength high harmonics. For HHG in gases, the ionization of the material enables phase matching through the free-electron contribution to the refractive index. In contrast, solid materials cannot tolerate significant ionization without permanent damage, so any ionization-based phase-matching scheme would not be practical. Fortunately, a key advantage of solid-state materials is that well-established techniques are available for periodically modifying the structure on the micrometer scale in order to achieve quasi-phase matching (QPM), which can allow for high conversion efficiencies, even in the absence of true phase matching. For example, the domain-reversal periodic poling of ferroelectric materials, such as lithium niobate, enables high-efficiency QPM for $\chi^{(2)}$ nonlinear processes such as second harmonic generation (SHG) \cite{fejer_quasi-phase-matched_1992}. Additionally, solid materials can exhibit large nonlinear susceptibilities, and they can be fabricated into waveguides with micrometer-scale cross-sections, tremendously enhancing frequency-conversion efficiencies in nonlinear processes.

Here we demonstrate that a high-confinement waveguide of chirped periodically poled lithium niobate (PPLN) can allow for the phase-matched generation of harmonics up to the 13$^\mathrm{th}$ order when pumped with 10~nJ pulses at 4100~nm. By using waveguides with different poling periods, we can control the output spectrum, indicating that QPM is playing an important role in the harmonic generation process. The total conversion efficiency of the harmonics (from the mid-infrared to the visible/ultraviolet region) is as high as 10\%, or about 1\% per harmonic in some cases. The high conversion efficiencies, low threshold pulse energies, and strong influence of quasi-phase-matching suggest that periodically poled waveguides are a promising platform for HHG at high repetition rates.

%%% SECTION: Experiment
\section{Experiment}

Ultrashort ($\sim$200~fs) pulses of 4100-nm light at a 1~MHz repetition rate are generated with a commercial laser system (KMLabs Y-Fi\textsuperscript{TM} OPA)~\cite{domingue_coherently_2017}. Since the 4100-nm light is generated via a white-light-seeded difference frequency generation process, the pulses (and any generated harmonics) should have stable carrier-envelope phase \cite{zimmermann_optical_2004}. The pulse energy delivered to the waveguide input is varied between 0.01~nJ and 50~nJ and input coupled using a chalcogenide aspheric lens. When the light is output-coupled using an identical lens, the total transmission through the PPLN is approximately 20\%. We make the assumption that the propagation loss through the waveguide is negligible, and that the coupling loss can be evenly divided between the input and output facets, providing an estimated input/output coupling efficiency of approximately 40\% per facet. Thus, we scale the measured incident laser power by 0.4 and discuss the results in terms of the pulse energy propagating in the waveguide. We note that the coupling loss experienced in this study is likely a result of un-optimized coupling optics, and coupling to PPLN waveguides with $>$80\% efficiency (per facet) has been experimentally demonstrated \cite{pelc_frequency_2012, nishikawa_efficient_2009}.

% Figure: PPLN
\begin{figure}[tb]
	\includegraphics[width=\linewidth]{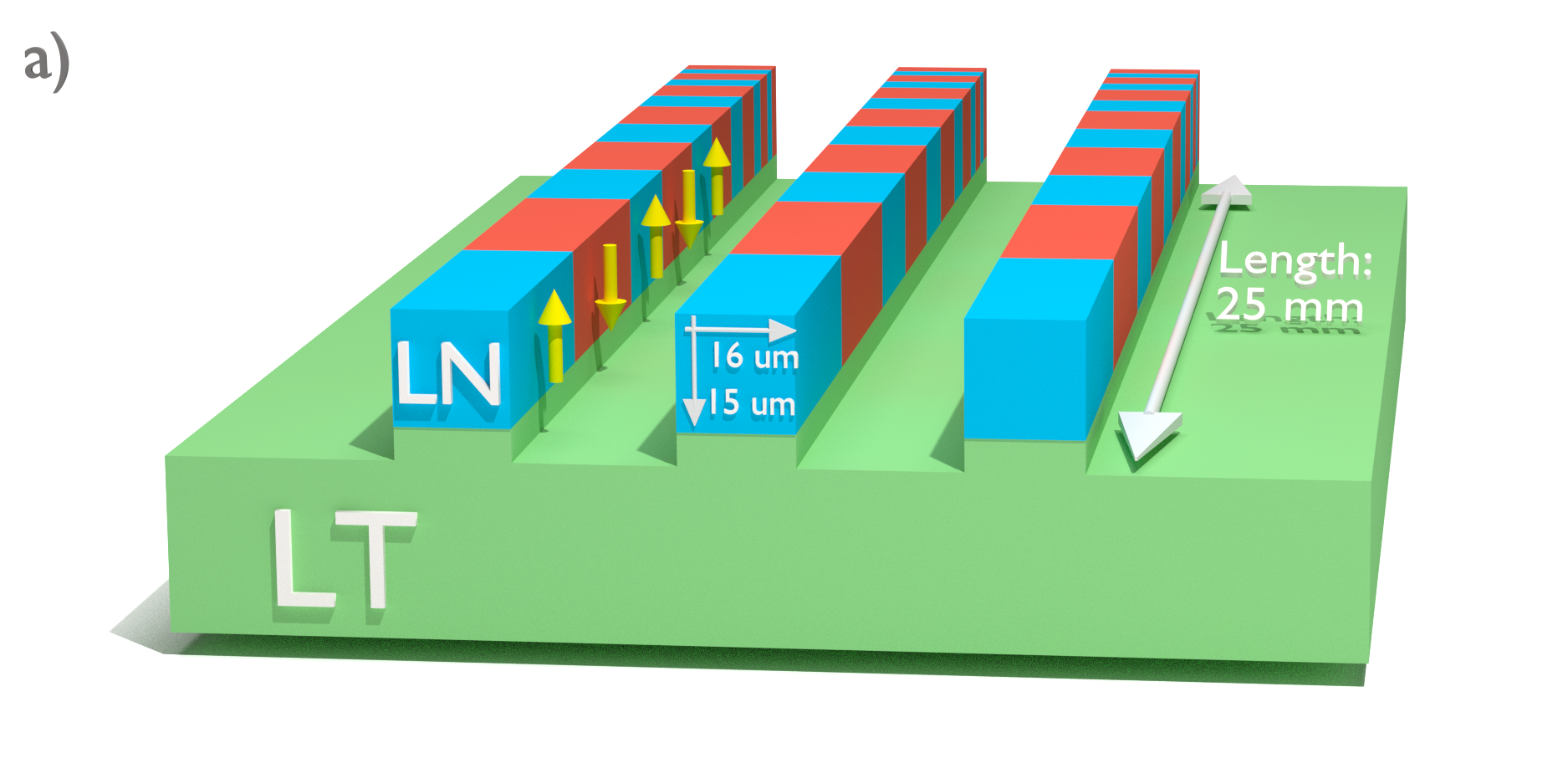}
    \includegraphics[width=\linewidth]{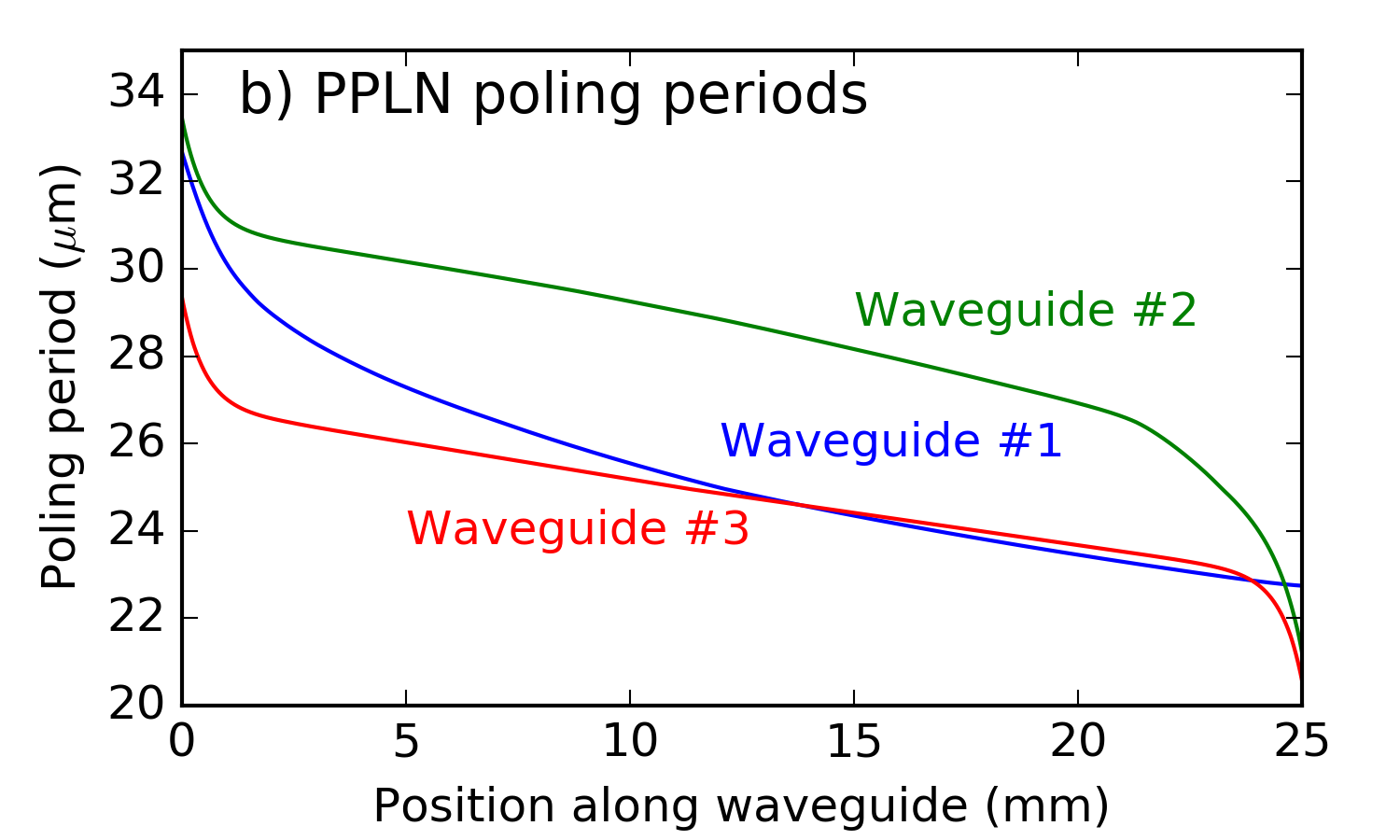}
	\caption{\label{fig:ppln} a) An illustration of the chip containing the lithium niobate (LN) waveguides on a lithium tantalate (LT) substrate. The pitch of the domain inversion periodic poling (shown as red and blue regions) changes along the length of the waveguide. The poling period is exaggerated for clarity; the waveguides contain several hundred poling periods. b) Each waveguide has a different rate of chirp for the poling period.}
\end{figure}

For broadband characterization of the output spectrum, the light emerging from the waveguide is collected with an $\mathrm{InF_3}$ multimode fiber and recorded using two optical spectrum analyzers (OSAs); a scanning-grating-based OSA (Ando AQ6315) is used to record the spectrum from 350 to 1700~nm, while a Fourier-transform OSA (Thorlabs OSA205) measures from 1000 to 5500~nm. In order to confirm the generation of short-wavelength light, we use a different output-coupling configuration, where a UV-fused silica lens is used to collimate the light, which is then coarsely dispersed with a calcium fluoride prism (to select only the ultraviolet wavelengths) and directed to a grating spectrometer (Ocean Optics Flame UV-Vis). While the infrared and visible spectrum analyzers provide measurements of the absolute power density, the UV spectrometer provides a spectrum with no power calibration, and we have rescaled it to match the visible OSA in the 800 THz region (Fig.~\ref{fig:overview}c). The power density recorded by the spectrum analyzers has been scaled by 1/0.4 to account for the estimated output-coupling loss.

The PPLN waveguides have a cross section of 15$\times$16~\si{\micro\meter} and a length of 25~mm (Fig.~\ref{fig:ppln}a), and were fabricated by NTT Electronics America. These ``direct-bonded'' ZnO-doped ridge-waveguides have a high resistance to photorefractive damage~\cite{nishida_direct-bonded_2003}. Each waveguide has a different function that determines the chirp of the poling period, which monotonically decreases along the length of the waveguide (Fig.~\ref{fig:ppln}b). The input laser polarization is set to vertical, which is the direction where the PPLN waveguides exhibit high $\chi^{(2)}$. Harmonics are generated by propagating the pump laser through each waveguide in both the ``increasing-poling-period'' and the ``decreasing-poling-period'' directions.

%%% SECTION: Results:
\section{Results}

Using 20 nJ of pump pulse energy, we see light generated across the visible and ultraviolet regions (Fig.~\ref{fig:overview}). This light is easily observable by eye, and discrete harmonics can be seen when the light is dispersed with a prism and visualized on a piece of white paper (Fig.~\ref{fig:overview}b). The ultraviolet harmonics can be seen as a faint blue glow on the right side of Fig.~\ref{fig:overview}b due to the fluorescence of the paper. Using a spectrometer to record the output light, harmonics can be seen across the entire visible region and into the ultraviolet region, up to $\mathrm{13^{th}}$ harmonic (H13, Fig.~\ref{fig:overview}c) at 315~nm. Because the bandgap of lithium niobate is near 310~nm, harmonics higher than H13 will only propagate through the lithium niobate for a short distance due to material absorption. It is possible that these shorter wavelengths are emitted, but with much lower flux than the below-bandgap harmonics, making detection challenging. When 20~mW of light is coupled into the waveguide, over 2~mW of light in the 350--1700~nm region is emitted, corresponding to a conversion efficiency from the mid-infrared to the visible of more than 10\%.

% Figure: power_scan
\begin{figure}
	\includegraphics[width=\linewidth]{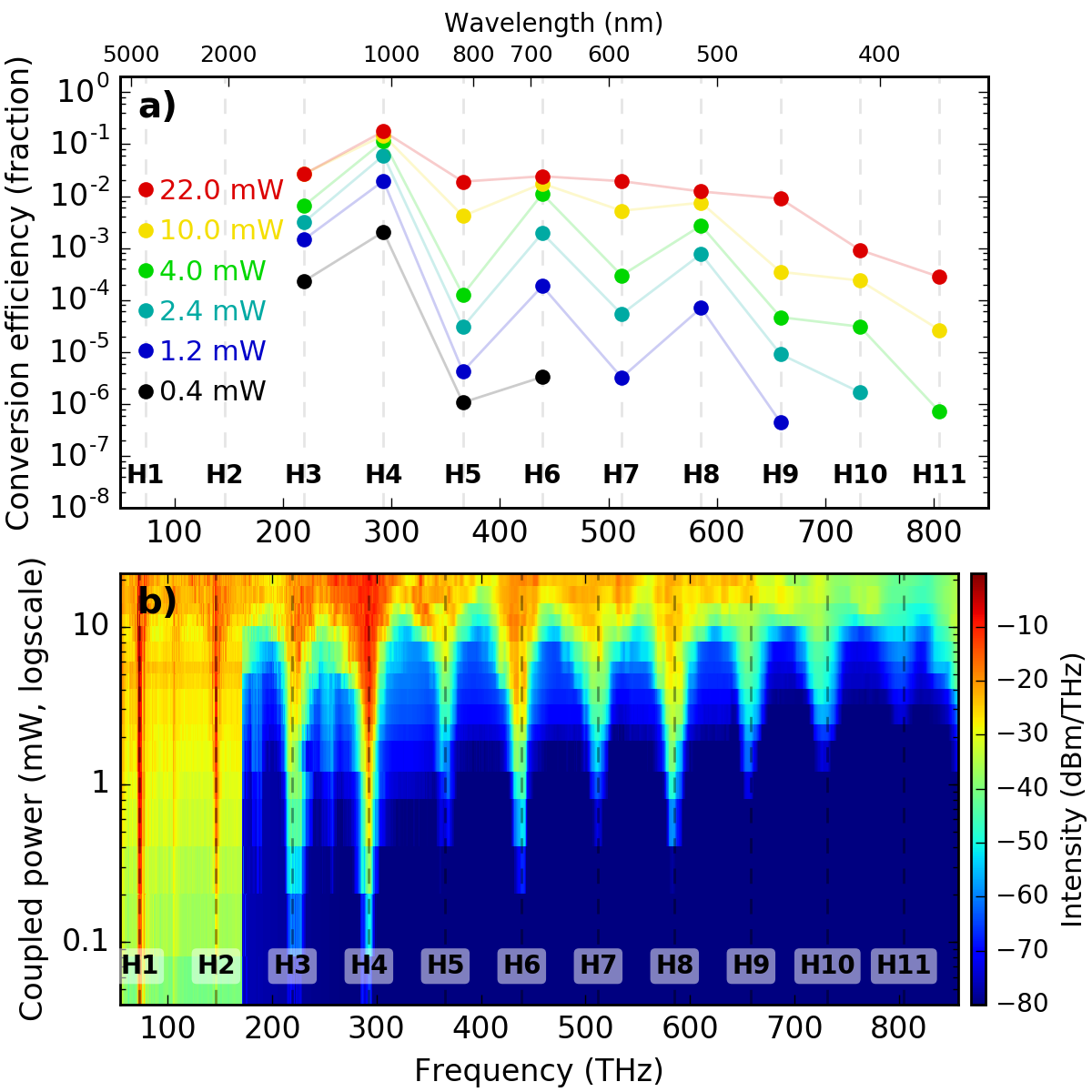}
	\caption{\label{fig:power} The conversion efficiency of the harmonics (a) and the output spectrum (b) at various pump power levels. For these results, Waveguide \#2 was pumped in the increasing-poling-period direction. (Note: Below 170 THz, the spectra were collected using a mid-infrared optical spectrum analyzer which has a higher noise-floor.)}
\end{figure}

The conversion efficiency to each harmonic order increases with increasing pump power (Fig.~\ref{fig:power}a). At low incident powers, the conversion efficiency drops sharply with increasing harmonic order in a perturbative fashion. However, at higher pump powers, a non-perturbative behavior is seen, and the harmonics form a plateau where the conversion efficiencies for H5 though H9 are roughly constant at approximately 1\% per harmonic. This behavior is similar to what is typically seen for HHG in gases and solids \cite{mcpherson_studies_1987, ferray_multiple-harmonic_1988, ghimire_observation_2011}. The spectral shape of the generated harmonics also undergoes a dramatic change with increasing pump power. At low pump powers, the spectrum consists of well-separated harmonics, while at higher pump powers of more than about 10~mW, the harmonics merge into a supercontinuum (Fig.~\ref{fig:power}b). It is possible that this coincides with a temporal compression of the pump pulse, as is typical of the soliton fission process that occurs in nonlinear media with anomalous dispersion \cite{dudley_supercontinuum_2006}. 

% Should we mention that since the continuum also forms around the 2f region near 1.9--2 microns, that normal-dispersion phase-mismatched SCG might also be taking place with this pulse? Or just cite some papers that might suggest that? -AK
% We mention that possibility at the end of the discussion section. I'm not sure if we should complicate the discussion here. -DH
 
% Figure: power_scaling
\begin{figure}
	\includegraphics[width=\linewidth]{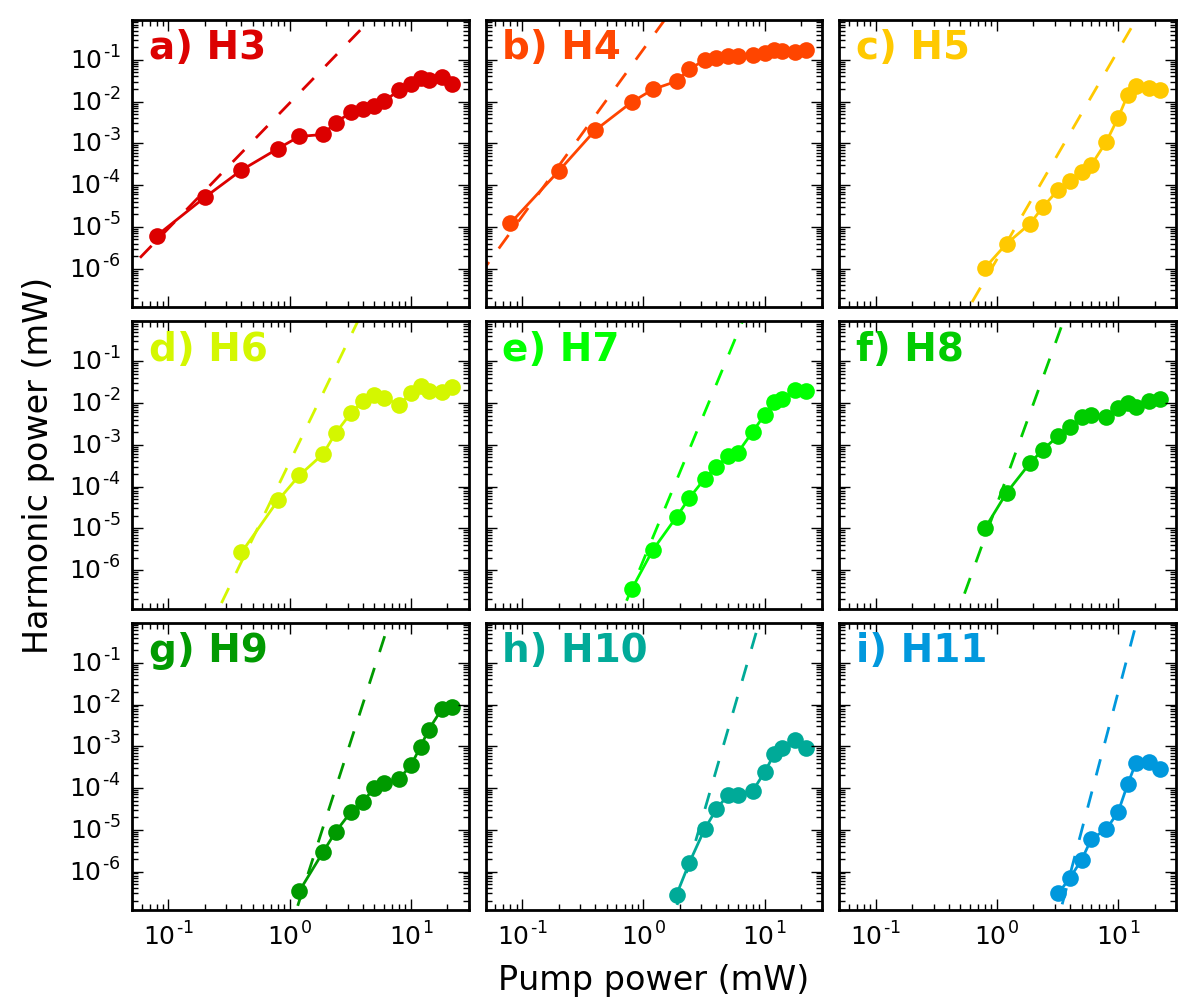}
	\caption{\label{fig:scaling} The integrated power of each harmonic order as a function of the pump power. At lower powers, the harmonics show a perturbative behavior, scaling roughly with $p^n$ (dashed lines), where $p$ is the pump power and $n$ is the harmonic order. At higher powers, a non-perturbative scaling is observed. For H9, H10, and H11, a step is seen at approximately 10~mW, which corresponds to the situation where the discrete harmonics transform into a supercontinuum.}
\end{figure}
% in panel b, make labels easier to read

The power in each harmonic order scales nonlinearly with pump power (Fig.~\ref{fig:scaling}), and each harmonic order exhibits a different power scaling. At low pump powers, the power scaling is close to $p^{n}$, where $p$ is the pump power and $n$ is the harmonic order. As the pump power increases, the exponent becomes smaller, which is also typical of HHG in both gases and solids as the power is increased beyond the perturbative regime \cite{ghimire_generation_2012, li_multiple-harmonic_1989,mcpherson_studies_1987, ferray_multiple-harmonic_1988}. Interestingly, for H7, H9, H10, and H11, a ``step'' is seen at around 10~mW of pump power. This sudden increase in conversion efficiency coincides with the spectral transition from separated harmonics to a supercontinuum, and provides further evidence that the peak intensity of the pump pulse may have been increased through temporal compression.

% Figure: low_power
\begin{figure}[tb]
	\includegraphics[width=\linewidth]{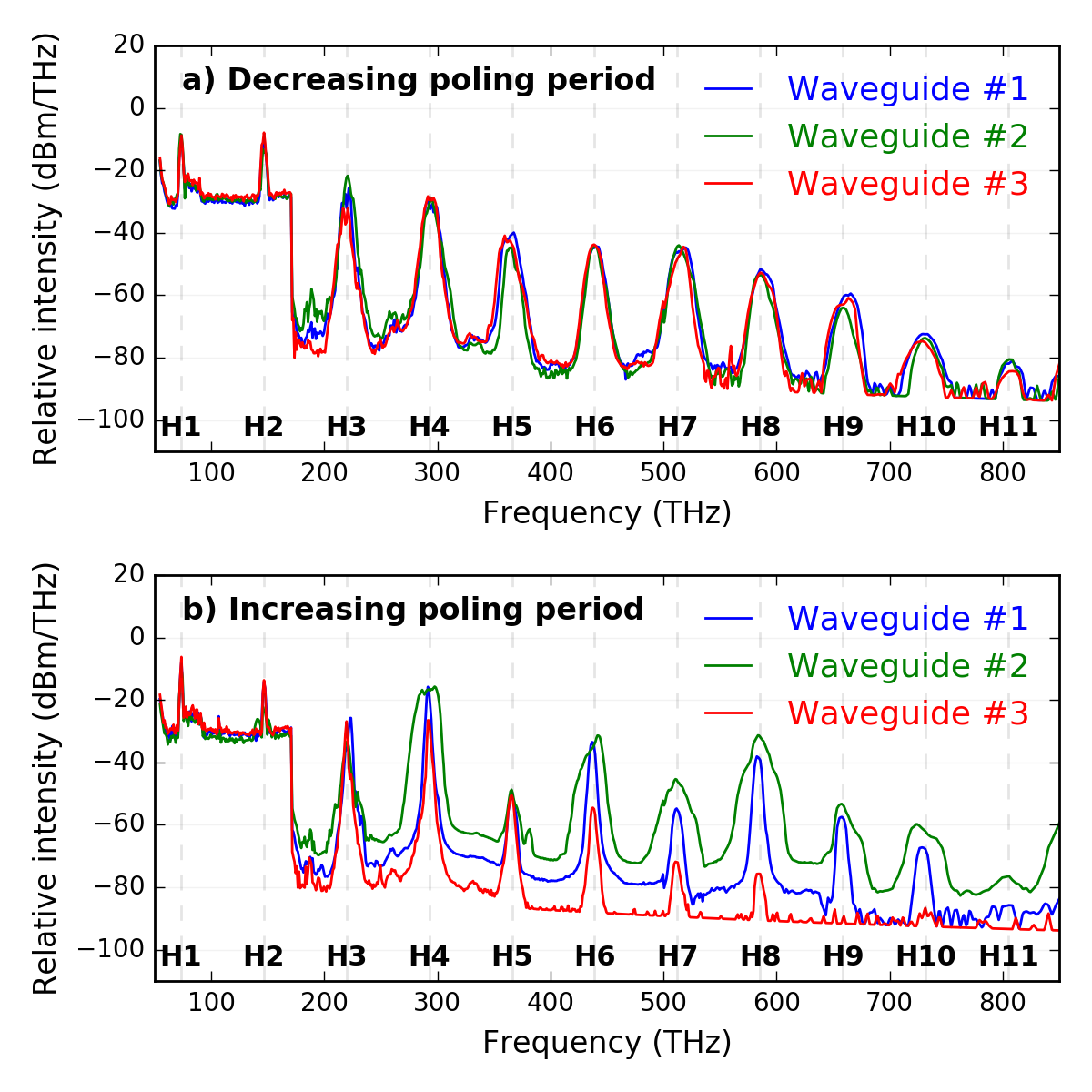}
	\caption{\label{fig:low_power} PPLN waveguides pumped with 5~nJ pulses. a) When pumped in the decreasing-poling-period direction, a smooth spectrum of even and odd harmonics is seen, with minimal changes between the waveguides. b) When pumped in the increasing-poling-period direction, significant differences are seen between the different waveguides.}
\end{figure}

% Figure: high_power
\begin{figure}[htb]
	\includegraphics[width=\linewidth]{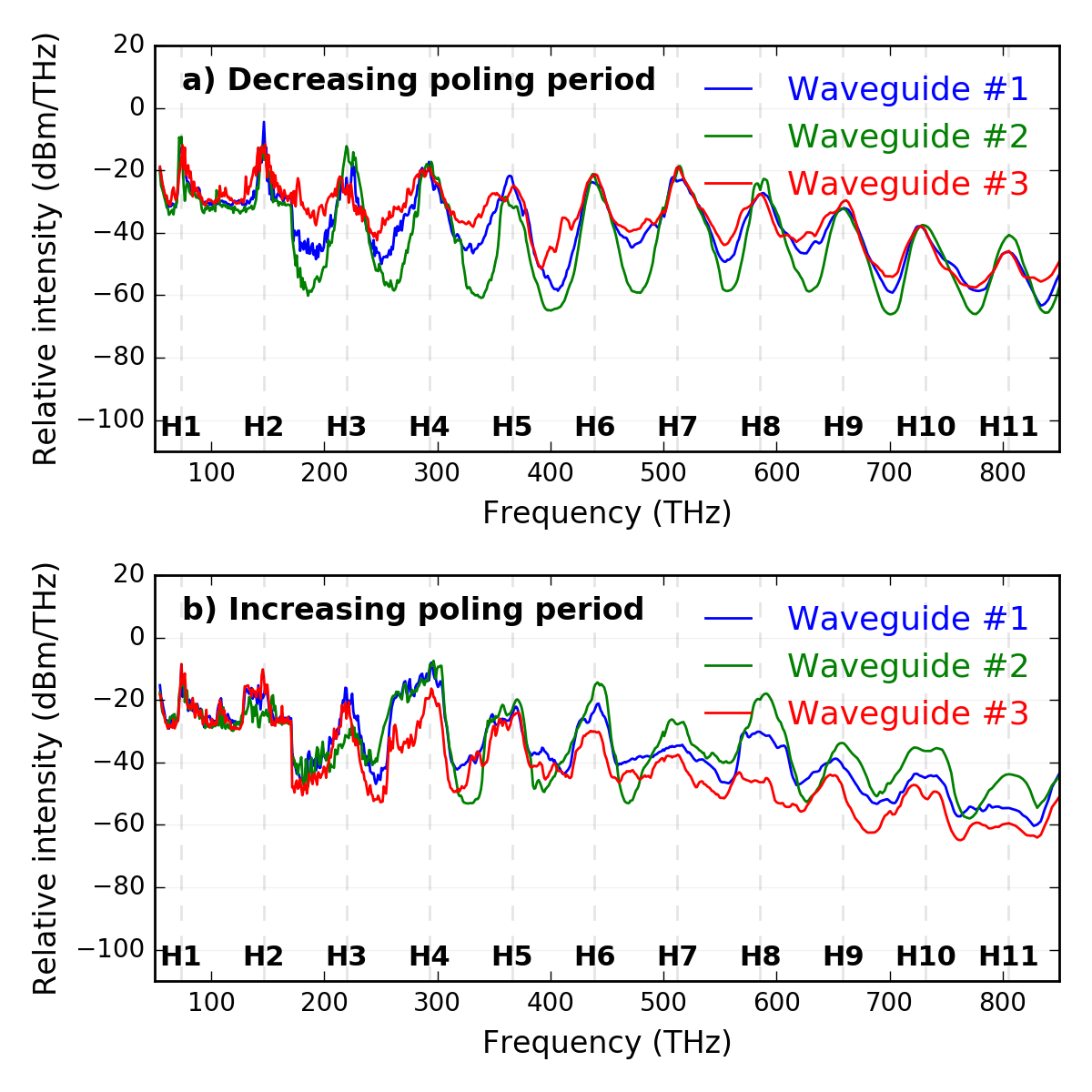}
	\caption{\label{fig:high_power} PPLN waveguides pumped with 25~nJ pulses exhibit broader harmonic peaks. a) When pumped in the decreasing-poling-period direction, small differences are seen between the various waveguides, mainly in the regions between the harmonic peaks. b) When pumped in the increasing-poling-period direction, large differences are seen in the spectra produced from the various waveguides.} 
\end{figure}

In order to identify the role of the periodic poling on harmonic generation process, we test three different waveguides, each with a different function determining the chirp of the poling period. Additionally, we pump each waveguide in both the increasing- and deceasing-poling-period directions. At low pump power (5~nJ, Fig.~\ref{fig:low_power}), we see an interesting effect: when pumping in the decreasing-poling-period direction, the harmonic spectra look similar for all of the waveguides (Fig.~\ref{fig:low_power}a). However, when pumped in the increasing-poling-period direction, each waveguide produces a different spectral shape (Fig.~\ref{fig:low_power}b). This large difference in the spectra generated by waveguides with different poling periods (Fig.~\ref{fig:low_power}b) is a clear indication that the periodic poling of the PPLN is providing QPM for the harmonic generation process. At higher powers (Fig.~\ref{fig:high_power}), the harmonic peaks broaden, and a supercontinuum-like spectrum is seen. Additionally, when the waveguides are pumped in the increasing poling period direction, the even-order harmonics are brighter than the odd-order harmonics. 

% SECTION: Discussion
\section{Discussion}
Having observed high conversion efficiency of mid-infrared light into visible and ultraviolet harmonics, and having experimentally confirmed that QPM is playing a significant role, the primary question is -- by what mechanism is the light generated? The currently available data do not unambiguously specify the mechanisms leading to the harmonic generation, but they do suggest three possibilities: 

\subsubsection*{Mechanism 1: Cascaded SHG and SFG}
\begin{figure}
	\includegraphics[width=\linewidth]{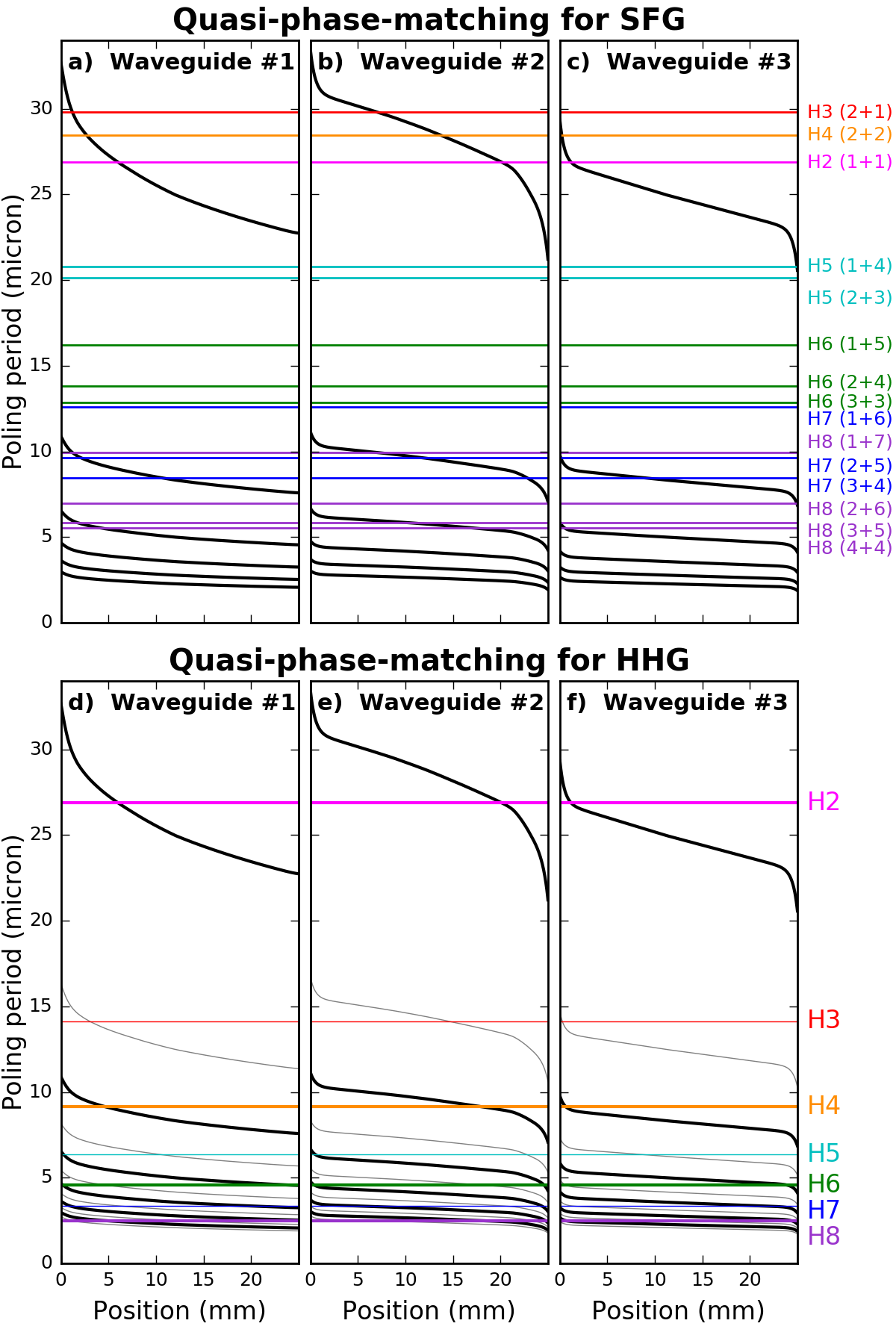} 
	\caption{\label{fig:phase_matching} a-c) The top black curve shows the poling period of the chirped PPLN waveguides, while lower curves indicate the effective poling period achieved via 3rd-, 5th-, 7th-, 9th-, and 11th-order QPM. The horizontal lines indicate the poling period required for QPM of various sum-frequency generation (SFG) processes (\textbf{Mechanism 1}). The labels indicate the harmonic order produced, and (in parentheses) the harmonic order of the two photons consumed in the SFG process. If the line crosses a curve, QPM is achieved at this location. While many of the SFG processes experience QPM, none of the pathways for H6 are quasi-phase matched for any of the waveguides. d-f) The required poling period for QPM the direct-HHG processes are shown with horizontal lines. Odd grating orders (thick black curves) can provide QPM for even harmonics (thick lines), while even grating orders (thin gray curves) can provide QPM for odd order harmonics (thin lines). In direct HHG (\textbf{Mechanism 2}) only odd grating orders (thick curves) are allowed, providing QPM only for even harmonics. In \textbf{Mechanism 3} both odd and even grating orders are allowed, providing QPM for all harmonic orders in all of the waveguides.}
\end{figure}

In this mechanism, the light is generated through the $\chi^{(2)}$ processes of SHG and sum-frequency generation (SFG). Recently, harmonics up to 8$^\mathrm{th}$ order were observed in a bulk chirped PPLN crystal~\cite{chen_high-efficiency_2015} and were attributed to cascaded $\chi^{(2)}$ processes, where each step of the process was separately quasi-phase-matched through the first and higher-order grating effects. However, in this case, this mechanism fails to explain the relatively smooth harmonic spectrum generated in our experiment. In particular, while most of the SFG processes experience QPM, SFG processes for H5 cannot be quasi-phase-matched in Waveguide \#1, and that SFG processes for H6 cannot be quasi-phase-matched in any of the waveguides (Fig.~\ref{fig:phase_matching}a-c). This conflicts with with our experimental data, which shows reasonable conversion to H5 and H6 for all of the waveguides (Figs.~\ref{fig:low_power} and \ref{fig:high_power}).

% Figure: phase_matching

\subsubsection*{Mechanism 2: Direct HHG}
In this mechanism, the 2nd, 3rd, 4th, 5th, etc. harmonics are generated directly from the $\chi^{(2)}$, $\chi^{(3)}$, $\chi^{(4)}$, $\chi^{(5)}$, etc. of the lithium niobate. This explanation is consistent with the smooth spectrum of even and odd harmonics that results from pumping in the decreasing-poling-period direction, which appears very similar to the spectrum generated from phase-mismatched HHG in non-centrosymmetric solids, such as ZnO \cite{ghimire_observation_2011, gholam-mirzaei_high_2017} and ZnSe \cite{lanin_mapping_2017}. The domain-reversal QPM in PPLN only allows phase-matching of even harmonics, since the odd-order nonlinearities ($\chi^{(3)}$, $\chi^{(5)}$, etc.) are symmetric and are consequently not affected by flipping the crystal axis. Thus, the presence of odd harmonics suggests that the light generated in the decreasing-poling-period direction is phase mismatched, while the light generated in the increasing-poling-period direction has an enhancement of the even harmonics (Fig.~\ref{fig:high_power}) due to QPM. Since the phase mismatch between the short-wavelength harmonics and the long-wavelength fundamental is quite large, very short poling periods would be required to achieve phase matching via the first order of the grating. However, in this case, QPM can be achieved for all even harmonic orders via higher-order grating effects (Fig.~\ref{fig:phase_matching}d-f), albeit with somewhat reduced efficiency. It is interesting to note that QPM for HHG in gases has been achieved using various schemes, including counter-propagating laser light \cite{zhang_quasi-phase-matching_2007, cohen_grating-assisted_2007, sidorenko_sawtooth_2010} and width-modulated gas-filled waveguides \cite{gibson_coherent_2003}, but has not seen widespread adoption, because other phase-matching techniques are available. To our knowledge, QPM of direct HHG using periodic poling has not been previously observed.

The peak intensities used in this study are consistent with a mechanism of direct HHG. For a 20~nJ, 200-fs (full-width at half-maximum) pulse, the peak intensity in the waveguide will be approximately $1.3\times10^{11}$~$\mathrm{W/cm^2}$. This is significantly less intense than the $10^{14}$~$\mathrm{W/cm^2}$ typically used for gas-phase HHG \cite{durfee_phase_1999}, or the $10^{13}$~$\mathrm{W/cm^2}$ used for generating HHG in relatively low-refractive-index solids like MgO \cite{you_high-harmonic_2017}, ZnO \cite{ghimire_observation_2011}, and solid argon \cite{ndabashimiye_solid-state_2016}. However, it is comparable to the peak intensities used to generate HHG in ZnSe \cite{lanin_mapping_2017}, a material with a higher refractive index. Previous work \cite{ettoumi_generalized_2010} has predicted that higher-order nonlinear susceptibilities should tend to increase with increasing refractive index, suggesting that higher index materials (like lithium niobate) may allow for efficient HHG with lower peak intensities.

While the direct-HHG mechanism explains many of the experimental observations, some things are left unexplained. First, it is not clear from the phase-matching analysis in Fig.~\ref{fig:phase_matching} why QPM should change significantly based on the propagation direction. Also, this mechanism implies that the generation of both even and odd harmonics when pumping in the decreasing-poling-period direction corresponds to phase-mismatched HHG, which conflicts with the excellent conversion efficiency seen in the experiment. 

\subsubsection*{Mechanism 3: Cascaded $\chi^{(2)}$ with phase-mismatched steps}
This mechanism is similar to Mechanism 1 in that it relies on cascaded SHG and SFG processes. The difference is that generation of intermediate harmonic orders is allowed to be phase mismatched, and the QPM conditions are calculated using the same equation as for direct high-harmonic generation. Indeed, several groups \cite{saltiel_direct_2006, das_direct_2006, delagnes_third_2011} have studied third harmonic generation via cascaded SHG and SFG in situations where both the SHG and SFG processes are, by themselves, phase mismatched, but the third harmonic process is phase matched, and they reported high conversion efficiencies (up to 25\% in the case of Ref.~\cite{saltiel_direct_2006}). Importantly, since the process is fundamentally based on cascaded $\chi^{(2)}$ processes, both even and odd harmonic orders can be quasi-phase matched by the domain-reversal periodic poling in a PPLN waveguide. While the \textit{even} harmonics experience typical QPM via the \textit{odd} order grating effects, QPM for the \textit{odd} harmonics is achieved via the \textit{even} order grating effects (Fig.~\ref{fig:phase_matching}d-f). Thus, the smooth spectrum of both even- and odd-order harmonics observed during pumping in the decreasing-poling-period direction (Fig.~\ref{fig:low_power}b) could be phase matched according to this mechanism. Additionally, since the conversion efficiency to each harmonic order is still influenced by the population of lower-order harmonics \cite{saltiel_direct_2006, das_direct_2006}, the direction of light propagation can affect the yield of all harmonic orders. However, this mechanism does not explain the preference for even-order harmonics in the case of pumping in the increasing-poling-period direction. 

\subsection*{Summary}
While each of the proposed mechanisms provides a reasonable explanation for some aspects of the results, none of the proposed mechanisms by itself provides a perfect explanation for all of the experimental observations, suggesting that some combination of these mechanisms or other nonlinear effects are at play. For example, at higher pump powers, the pump spectrum broadens, and it's likely that significant self-phase-modulation and pulse compression are occurring, which could increase the peak intensity of the pulse as it propagates along the waveguide. Additionally, phase-mismatched cascaded $\chi^{(2)}$ processes in PPLN waveguides are known to provide an alternative pathway for processes such as self-phase modulation and soliton fission \cite{bache_scaling_2007, moses_soliton_2006}. Finally, light could be generated into higher-order spatial modes of the waveguide \cite{lee_mode_2010, ming_high_2005, park_spatial_2014, wang_second_2017}, further complicating the analysis. More sophisticated modeling will be required in order to definitively explain how these harmonics are generated and predict how they can be optimized in future experiments.

\subsection*{Outlook}
While the results here focus on the high-efficiency generation of visible and ultraviolet light using nJ-level pulses at a 1~MHz repetition rate, the technique of HHG in periodically poled waveguides can likely be scaled into other regimes. For example, waveguides with smaller cross sections would enable even lower pulse energies and higher repetition rates. Additionally, shorter poling periods would allow for the compensation of higher phase-mismatch, and could enable the QPM of HHG in the extreme ultraviolet or x-ray regions. Lastly, more careful control over the waveguide dispersion and pump propagation could provide temporal compression of pump pulses to the few-cycle scale \cite{dudley_supercontinuum_2006}. All together, periodically poled waveguides could provide an ideal combination of temporal confinement, spatial confinement, high nonlinearity, and flexible phase-matching, which could support HHG at high repetition rates with ultracompact lasers.

\section{Conclusion}
In summary, have demonstrated broadband high-harmonic generation (up to the 13$\mathrm{^{th}}$ harmonic order) using nJ-level pulses at 1-MHz, with total conversion efficiencies as high as 10\%. The high conversion efficiency and high repetition rate are enabled through the use of periodically poled waveguides. The waveguide geometry provides high spatial confinement of the light over a long interaction region, while the periodic poling provides quasi-phase-matching, greatly enhancing the efficiency. The harmonic spectrum and power scaling exhibit the non-perturbative behavior typical of the direct $\chi^{(n)}$ high-harmonic generation process, while the apparent phase matching of both odd- and even-order harmonics, as well as the dependence of the spectrum on the direction of the poling-period chirp, suggest a mechanism based on cascaded $\chi^{(2)}$ effects. Regardless of the precise mechanism, this ability to generate high-order harmonics with excellent conversion efficiency using low pulse energies could enable compact, solid-state, high-repetition-rate sources of ultrashort pulses \cite{hassan_optical_2016} and provide new methods of probing the electronic structure of solid-state materials.

% Acknowledgements
\section*{Funding information}
This material is based upon work supported by the Air Force Office of Scientific Research under award number FA9550-16-1-0016, 
the Defense Advanced Research Projects Agency (DARPA) ACES, PULSE, and SCOUT programs, 
the National Aeronautics and Space Administration (NASA), 
the National Institute of Standards and Technology (NIST), 
and the National Research Council (NRC). KMLabs researchers acknowledge support from the DARPA PULSE program.

\section*{Acknowledgements}
We acknowledge helpful discussions with Kevin Dorney, Jennifer Ellis, Shambhu Ghimire, Carlos Hernandez-Garcia, and Sterling Backus. We also thank the reviewers from the NIST Boulder Editorial Review Board, including Ian Coddington, Norman Sanford, Chris Oates, and John Lowe, for providing useful feedback on this manuscript.

\subsection*{Disclaimer}
Note: certain commercial equipment, instruments, or materials are identified in this paper in order to specify the experimental procedure adequately. Such identification is not intended to imply recommendation or endorsement by the National Institute of Standards and Technology, nor is it intended to imply that the materials or equipment identified are necessarily the best available for the purpose. This work is a contribution of the United States government and is not subject to copyright in the United States of America.

\bibliography{Zotero}

\begin{thebibliography}{10}
\newcommand{\enquote}[1]{``#1''}

\bibitem{zhao_tailoring_2012}
K.~Zhao, Q.~Zhang, M.~Chini, Y.~Wu, X.~Wang, and Z.~Chang, \enquote{Tailoring a
  67 attosecond pulse through advantageous phase-mismatch,} Optics Letters
  \textbf{37}, 3891--3893 (2012).

\bibitem{sandberg_lensless_2007}
R.~L. Sandberg, A.~Paul, D.~A. Raymondson, S.~Hädrich, D.~M. Gaudiosi,
  J.~Holtsnider, R.~I. Tobey, O.~Cohen, M.~M. Murnane, H.~C. Kapteyn, C.~Song,
  J.~Miao, Y.~Liu, and F.~Salmassi, \enquote{Lensless {Diffractive} {Imaging}
  {Using} {Tabletop} {Coherent} {High}-{Harmonic} {Soft}-{X}-{Ray} {Beams},}
  Physical Review Letters \textbf{99}, 098103 (2007).

\bibitem{gardner_subwavelength_2017}
D.~F. Gardner, M.~Tanksalvala, E.~R. Shanblatt, X.~Zhang, B.~R. Galloway, C.~L.
  Porter, R.~Karl~Jr, C.~Bevis, D.~E. Adams, H.~C. Kapteyn, M.~M. Murnane, and
  G.~F. Mancini, \enquote{Subwavelength coherent imaging of periodic samples
  using a 13.5 nm tabletop high-harmonic light source,} Nature Photonics
  \textbf{11}, 259--263 (2017).

\bibitem{zurch_real-time_2014}
M.~Zürch, J.~Rothhardt, S.~Hädrich, S.~Demmler, M.~Krebs, J.~Limpert,
  A.~Tünnermann, A.~Guggenmos, U.~Kleineberg, and C.~Spielmann,
  \enquote{Real-time and {Sub}-wavelength {Ultrafast} {Coherent} {Diffraction}
  {Imaging} in the {Extreme} {Ultraviolet},} Scientific Reports \textbf{4},
  7356 (2014).

\bibitem{leone_attosecond_2016}
S.~Leone and D.~Neumark, \enquote{Attosecond science in atomic, molecular, and
  condensed matter physics,} Faraday Discussions \textbf{194}, 15--39 (2016).

\bibitem{zhou_probing_2012}
X.~Zhou, P.~Ranitovic, C.~W. Hogle, J.~H.~D. Eland, H.~C. Kapteyn, and M.~M.
  Murnane, \enquote{Probing and controlling non-{Born}-{Oppenheimer} dynamics
  in highly excited molecular ions,} Nature Physics \textbf{8}, 232--237
  (2012).

\bibitem{kfir_nanoscale_2017}
O.~Kfir, S.~Zayko, C.~Nolte, M.~Sivis, M.~Möller, B.~Hebler, S.~S. P.~K.
  Arekapudi, D.~Steil, S.~Schäfer, M.~Albrecht, O.~Cohen, S.~Mathias, and
  C.~Ropers, \enquote{Nanoscale {Magnetic} {Imaging} using {Circularly}
  {Polarized} {High}-{Harmonic} {Radiation},} arXiv \textbf{1706.07695} (2017).

\bibitem{tao_direct_2016}
Z.~Tao, C.~Chen, T.~Szilvási, M.~Keller, M.~Mavrikakis, H.~Kapteyn, and
  M.~Murnane, \enquote{Direct time-domain observation of attosecond final-state
  lifetimes in photoemission from solids,} Science \textbf{353}, 62--67 (2016).

\bibitem{chen_distinguishing_2017}
C.~Chen, Z.~Tao, A.~Carr, P.~Matyba, T.~Szilvási, S.~Emmerich, M.~Piecuch,
  M.~Keller, D.~Zusin, S.~Eich, M.~Rollinger, W.~You, S.~Mathias, U.~Thumm,
  M.~Mavrikakis, M.~Aeschlimann, P.~M. Oppeneer, H.~Kapteyn, and M.~Murnane,
  \enquote{Distinguishing attosecond electron–electron scattering and
  screening in transition metals,} Proceedings of the National Academy of
  Sciences \textbf{114}, E5300--E5307 (2017).

\bibitem{ghimire_observation_2011}
S.~Ghimire, A.~D. DiChiara, E.~Sistrunk, P.~Agostini, L.~F. DiMauro, and D.~A.
  Reis, \enquote{Observation of high-order harmonic generation in a bulk
  crystal,} Nature Physics \textbf{7}, 138--141 (2011).

\bibitem{hohenleutner_real-time_2015}
M.~Hohenleutner, F.~Langer, O.~Schubert, M.~Knorr, U.~Huttner, S.~W. Koch,
  M.~Kira, and R.~Huber, \enquote{Real-time observation of interfering crystal
  electrons in high-harmonic generation,} Nature \textbf{523}, 572--575 (2015).

\bibitem{gholam-mirzaei_high_2017}
S.~Gholam-Mirzaei, J.~Beetar, and M.~Chini, \enquote{High harmonic generation
  in {ZnO} with a high-power mid-{IR} {OPA},} Applied Physics Letters
  \textbf{110}, 061101 (2017).

\bibitem{han_high-harmonic_2016}
S.~Han, H.~Kim, Y.~W. Kim, Y.-J. Kim, S.~Kim, I.-Y. Park, and S.-W. Kim,
  \enquote{High-harmonic generation by field enhanced femtosecond pulses in
  metal-sapphire nanostructure,} Nature Communications \textbf{7}, 13105
  (2016).

\bibitem{kim_generation_2017}
H.~Kim, S.~Han, Y.~W. Kim, S.~Kim, and S.-W. Kim, \enquote{Generation of
  {Coherent} {Extreme}-{Ultraviolet} {Radiation} from {Bulk} {Sapphire}
  {Crystal},} ACS Photonics \textbf{4}, 1627--1632 (2017).

\bibitem{luu_extreme_2015}
T.~T. Luu, M.~Garg, S.~Y. Kruchinin, A.~Moulet, M.~T. Hassan, and
  E.~Goulielmakis, \enquote{Extreme ultraviolet high-harmonic spectroscopy of
  solids,} Nature \textbf{521}, 498--502 (2015).

\bibitem{ndabashimiye_solid-state_2016}
G.~Ndabashimiye, S.~Ghimire, M.~Wu, D.~A. Browne, K.~J. Schafer, M.~B. Gaarde,
  and D.~A. Reis, \enquote{Solid-state harmonics beyond the atomic limit,}
  Nature \textbf{534}, 520--523 (2016).

\bibitem{schubert_sub-cycle_2014}
O.~Schubert, M.~Hohenleutner, F.~Langer, B.~Urbanek, C.~Lange, U.~Huttner,
  D.~Golde, T.~Meier, M.~Kira, S.~W. Koch, and R.~Huber, \enquote{Sub-cycle
  control of terahertz high-harmonic generation by dynamical {Bloch}
  oscillations,} Nature Photonics \textbf{8}, 119--123 (2014).

\bibitem{yoshikawa_high-harmonic_2017}
N.~Yoshikawa, T.~Tamaya, and K.~Tanaka, \enquote{High-harmonic generation in
  graphene enhanced by elliptically polarized light excitation,} Science
  \textbf{356}, 736--738 (2017).

\bibitem{vampa_all-optical_2015}
G.~Vampa, T.~J. Hammond, N.~Thiré, B.~E. Schmidt, F.~Légaré, C.~R. McDonald,
  T.~Brabec, D.~D. Klug, and P.~B. Corkum, \enquote{All-{Optical}
  {Reconstruction} of {Crystal} {Band} {Structure},} Physical Review Letters
  \textbf{115}, 193603 (2015).

\bibitem{lanin_mapping_2017}
A.~A. Lanin, E.~A. Stepanov, A.~B. Fedotov, and A.~M. Zheltikov,
  \enquote{Mapping the electron band structure by intraband high-harmonic
  generation in solids,} Optica \textbf{4}, 516--519 (2017).

\bibitem{you_anisotropic_2017}
Y.~S. You, D.~A. Reis, and S.~Ghimire, \enquote{Anisotropic high-harmonic
  generation in bulk crystals,} Nature Physics \textbf{13}, 345--349 (2017).

\bibitem{tancogne-dejean_impact_2017}
N.~Tancogne-Dejean, O.~D. Mücke, F.~X. Kärtner, and A.~Rubio, \enquote{Impact
  of the {Electronic} {Band} {Structure} in {High}-{Harmonic} {Generation}
  {Spectra} of {Solids},} Physical Review Letters \textbf{118}, 087403 (2017).

\bibitem{rundquist_phase-matched_1998}
A.~Rundquist, C.~G. Durfee, Z.~Chang, C.~Herne, S.~Backus, M.~M. Murnane, and
  H.~C. Kapteyn, \enquote{Phase-{Matched} {Generation} of {Coherent} {Soft}
  {X}-rays,} Science \textbf{280}, 1412--1415 (1998).

\bibitem{tamaki_highly_1999}
Y.~Tamaki, J.~Itatani, Y.~Nagata, M.~Obara, and K.~Midorikawa, \enquote{Highly
  {Efficient}, {Phase}-{Matched} {High}-{Harmonic} {Generation} by a
  {Self}-{Guided} {Laser} {Beam},} Physical Review Letters \textbf{82},
  1422--1425 (1999).

\bibitem{durfee_phase_1999}
C.~G. Durfee, A.~R. Rundquist, S.~Backus, C.~Herne, M.~M. Murnane, and H.~C.
  Kapteyn, \enquote{Phase {Matching} of {High}-{Order} {Harmonics} in {Hollow}
  {Waveguides},} Physical Review Letters \textbf{83}, 2187--2190 (1999).

\bibitem{zhang_quasi-phase-matching_2007}
X.~Zhang, A.~L. Lytle, T.~Popmintchev, X.~Zhou, H.~C. Kapteyn, M.~M. Murnane,
  and O.~Cohen, \enquote{Quasi-phase-matching and quantum-path control of
  high-harmonic generation using counterpropagating light,} Nature Physics
  \textbf{3}, 270--275 (2007).

\bibitem{cohen_grating-assisted_2007}
O.~Cohen, X.~Zhang, A.~L. Lytle, T.~Popmintchev, M.~M. Murnane, and H.~C.
  Kapteyn, \enquote{Grating-{Assisted} {Phase} {Matching} in {Extreme}
  {Nonlinear} {Optics},} Physical Review Letters \textbf{99}, 053902 (2007).

\bibitem{ghimire_generation_2012}
S.~Ghimire, A.~D. DiChiara, E.~Sistrunk, G.~Ndabashimiye, U.~B. Szafruga,
  A.~Mohammad, P.~Agostini, L.~F. DiMauro, and D.~A. Reis, \enquote{Generation
  and propagation of high-order harmonics in crystals,} Physical Review A
  \textbf{85}, 043836 (2012).

\bibitem{fejer_quasi-phase-matched_1992}
M.~M. Fejer, G.~A. Magel, D.~H. Jundt, and R.~L. Byer,
  \enquote{Quasi-phase-matched second harmonic generation: tuning and
  tolerances,} IEEE Journal of Quantum Electronics \textbf{28}, 2631--2654
  (1992).

\bibitem{domingue_coherently_2017}
S.~Domingue, D.~G. Winters, M.~Kirchner, S.~J. Backus, and S.~J. Backus,
  \enquote{Coherently seeded optical parametric amplifier with 500 {nJ}
  short-wave infrared signal at 1 {MHz},} in \enquote{Conference on {Lasers}
  and {Electro}-{Optics} (2017), paper {JTh}2A.129,}  (Optical Society of
  America, 2017), p. JTh2A.129.

\bibitem{zimmermann_optical_2004}
M.~Zimmermann, C.~Gohle, R.~Holzwarth, T.~Udem, and T.~W. Hänsch,
  \enquote{Optical clockwork with an offset-free difference-frequency comb:
  accuracy of sum- and difference-frequency generation,} Optics Letters
  \textbf{29}, 310--312 (2004).

\bibitem{pelc_frequency_2012}
J.~Pelc, \enquote{Frequency {Conversion} of {Single} {Photons}: {Physics},
  {Devices}, and {Applications},} Ph.D. thesis, Stanford University (2012).

\bibitem{nishikawa_efficient_2009}
T.~Nishikawa, A.~Ozawa, Y.~Nishida, M.~Asobe, F.-L. Hong, and T.~W. Hänsch,
  \enquote{Efficient 494 {mW} sum-frequency generation of sodium resonance
  radiation at 589 nm by using a periodically poled {Zn}:{LiNbO}$_{\textrm{3}}$
  ridge waveguide,} Optics Express \textbf{17}, 17792--17800 (2009).

\bibitem{nishida_direct-bonded_2003}
Y.~Nishida, H.~Miyazawa, M.~Asobe, O.~Tadanaga, and H.~Suzuki,
  \enquote{Direct-bonded {QPM}-{LN} ridge waveguide with high damage resistance
  at room temperature,} Electronics Letters \textbf{39}, 609--611 (2003).

\bibitem{mcpherson_studies_1987}
A.~McPherson, G.~Gibson, H.~Jara, U.~Johann, T.~S. Luk, I.~A. McIntyre,
  K.~Boyer, and C.~K. Rhodes, \enquote{Studies of multiphoton production of
  vacuum-ultraviolet radiation in the rare gases,} JOSA B \textbf{4}, 595--601
  (1987).

\bibitem{ferray_multiple-harmonic_1988}
M.~Ferray, A.~L'Huillier, X.~F. Li, L.~A. Lompre, G.~Mainfray, and C.~Manus,
  \enquote{Multiple-harmonic conversion of 1064 nm radiation in rare gases,}
  Journal of Physics B: Atomic, Molecular and Optical Physics \textbf{21}, L31
  (1988).

\bibitem{dudley_supercontinuum_2006}
J.~M. Dudley, G.~Genty, and S.~Coen, \enquote{Supercontinuum generation in
  photonic crystal fiber,} Reviews of Modern Physics \textbf{78}, 1135--1184
  (2006).

\bibitem{li_multiple-harmonic_1989}
X.~F. Li, A.~L’Huillier, M.~Ferray, L.~A. Lompré, and G.~Mainfray,
  \enquote{Multiple-harmonic generation in rare gases at high laser intensity,}
  Physical Review A \textbf{39}, 5751--5761 (1989).

\bibitem{chen_high-efficiency_2015}
B.-Q. Chen, C.~Zhang, C.-Y. Hu, R.-J. Liu, and Z.-Y. Li,
  \enquote{High-{Efficiency} {Broadband} {High}-{Harmonic} {Generation} from a
  {Single} {Quasi}-{Phase}-{Matching} {Nonlinear} {Crystal},} Phys. Rev. Lett.
  \textbf{115}, 083902 (2015).

\bibitem{sidorenko_sawtooth_2010}
P.~Sidorenko, M.~Kozlov, A.~Bahabad, T.~Popmintchev, M.~Murnane, H.~Kapteyn,
  and O.~Cohen, \enquote{Sawtooth grating-assisted phase-matching,} Optics
  Express \textbf{18}, 22686--22692 (2010).

\bibitem{gibson_coherent_2003}
E.~A. Gibson, A.~Paul, N.~Wagner, R.~Tobey, D.~Gaudiosi, S.~Backus, I.~P.
  Christov, A.~Aquila, E.~M. Gullikson, D.~T. Attwood, M.~M. Murnane, and H.~C.
  Kapteyn, \enquote{Coherent {Soft} {X}-ray {Generation} in the {Water}
  {Window} with {Quasi}-{Phase} {Matching},} Science \textbf{302}, 95--98
  (2003).

\bibitem{you_high-harmonic_2017}
Y.~S. You, Y.~Yin, Y.~Wu, A.~Chew, X.~Ren, F.~Zhuang, S.~Gholam-Mirzaei,
  M.~Chini, Z.~Chang, and S.~Ghimire, \enquote{High-harmonic generation in
  amorphous solids,} arXiv:1705.07854 [physics]  (2017). ArXiv: 1705.07854.

\bibitem{ettoumi_generalized_2010}
W.~Ettoumi, Y.~Petit, J.~Kasparian, and J.-P. Wolf, \enquote{Generalized
  {Miller} {Formulæ},} Optics Express \textbf{18}, 6613--6620 (2010).

\bibitem{saltiel_direct_2006}
S.~M. Saltiel, P.~K. Datta, K.~Koynov, and V.~L. Saltiel, \enquote{Direct
  {Third} {Harmonic} {Generation} in {Single} {Quadratic} {Crystal} in {Quasi}
  {Phase} {Matched} {Regime},} Bulg. J. Phys. \textbf{33}, 1310 (2006).

\bibitem{das_direct_2006}
S.~K. Das, S.~Mukhopadhyay, N.~Sinha, A.~Saha, P.~K. Datta, S.~M. Saltiel, and
  L.~C. Andreani, \enquote{Direct third harmonic generation due to quadratic
  cascaded processes in periodically poled crystals,} Optics Communications
  \textbf{262}, 108--113 (2006).

\bibitem{delagnes_third_2011}
J.~C. Delagnes and L.~Canioni, \enquote{Third harmonic generation in
  periodically poled crystals,} in \enquote{Nonlinear {Frequency} {Generation}
  and {Conversion}: {Materials}, {Devices}, and {Applications},} , vol. 7917
  (2011), vol. 7917, pp. 79171C--79171C--9.

\bibitem{bache_scaling_2007}
M.~Bache, J.~Moses, and F.~W. Wise, \enquote{Scaling laws for soliton pulse
  compression by cascaded quadratic nonlinearities,} JOSA B \textbf{24},
  2752--2762 (2007).

\bibitem{moses_soliton_2006}
J.~Moses and F.~W. Wise, \enquote{Soliton compression in quadratic media:
  high-energy few-cycle pulses with a frequency-doubling crystal,} Optics
  Letters \textbf{31}, 1881--1883 (2006).

\bibitem{lee_mode_2010}
Y.~L. Lee, W.~Shin, B.-A. Yu, C.~Jung, Y.-C. Noh, and D.-K. Ko, \enquote{Mode
  {Tailoring} in a {Ridge}-type {Periodically} {Poled} {Lithium} {Niobate}
  {Waveguide},} Optics Express \textbf{18}, 7678--7684 (2010).

\bibitem{ming_high_2005}
L.~Ming, C.~B.~E. Gawith, K.~Gallo, M.~V. O’Connor, G.~D. Emmerson, and
  P.~G.~R. Smith, \enquote{High conversion efficiency single-pass second
  harmonic generation in a zinc-diffused periodically poled lithium niobate
  waveguide,} Optics Express \textbf{13}, 4862--4868 (2005).

\bibitem{park_spatial_2014}
J.-H. Park, T.-Y. Kang, J.-H. Ha, and H.-Y. Lee, \enquote{Spatial mode behavior
  of second harmonic generation in a ridge-type waveguide with a periodically
  poled {MgO}-doped lithium niobate crystal,} Japanese Journal of Applied
  Physics \textbf{53}, 062201 (2014).

\bibitem{wang_second_2017}
C.~Wang, X.~Xiong, N.~Andrade, V.~Venkataraman, X.-F. Ren, G.-C. Guo, and
  M.~Lončar, \enquote{Second harmonic generation in nano-structured thin-film
  lithium niobate waveguides,} Optics Express \textbf{25}, 6963--6973 (2017).

\bibitem{hassan_optical_2016}
M.~T. Hassan, T.~T. Luu, A.~Moulet, O.~Raskazovskaya, P.~Zhokhov, M.~Garg,
  N.~Karpowicz, A.~M. Zheltikov, V.~Pervak, F.~Krausz, and E.~Goulielmakis,
  \enquote{Optical attosecond pulses and tracking the nonlinear response of
  bound electrons,} Nature \textbf{530}, 66--70 (2016).

\end{thebibliography}

\end{document}